\documentclass[12pt,preprint]{aastex}

\usepackage{subfigure} 

\shorttitle{Plume Geometrical Model}
\shortauthors{Peng}

\begin{document}


\title{A geometrical model for polar plumes observed by STEREO SECCHI}

\author{Beidi Peng}
\affil{Physics and Astronomy Department,  206 Gallalee Hall, 514 University Blvd. The University of Alabama, Tuscaloosa, AL 35487-0324}
\email{bpeng1@crimson.ua.edu}

\begin{abstract}
Solar polar plumes are bright radial rays rooted at the sun's polar areas. They are widely believed to have the structure of expanding tube.A four degree polynomial function was assumed to represent the change of cross section diameter as a function of height and four unknown parameters were calculated with measured average widths of total 31 plumes at 4 heights(1.04$R_s$,1.10$R_s$,1.16$R_s$,1.20$R_s$).
\end{abstract}

\keywords{sun: UV radiation --- sun: corona}

\section{Introduction}

Solar polar plumes are bright rays located at coronal holes or polar areas. They can have different heights depending on particular wavelength of the observation. Solar polar plumes are visiable from the base until approximately 1.2 $R_s$ if observed in ultraviolet wavelength.(eg. STEREO telescope EUVI).  Observations in x-rays(eg.{\it Hinode} XRT) are mainly only for hot gas distributed on the solar surface and bright points of the base of polar plumes. 

Because they locate at coronal hole regions where the fast solar wind is originated, solar polar plumes have been associated with a possible solar fast wind source. However, there still lacks evidence about this connection in terms of the underneath physical process.\citet{rao07} have shown plumes have higher electron density than interplumes' but approach that of interplume regions as increasing height.\citet{ gio97, koh95} have also shown the H I Ly$\alpha$ line has narrower width of plumes than that of interplumes,corresponding with a lower temperature. \citet{has97} has the same results as \citep{gio97} by analyzing different UV lines(O vi $\lambda$ 1032 line width is lower by 10\%-15\%).\citet{wil98,you99} have studied intensity ratios of UV spectral lines in the low altitude of the corona. They found an increasing of temperature with respect to heights in the background corona(interplumes),but a similar temperature for lower parts of plumes.

The geometry of polar plumes have been studied and argued for a long time. There are mainly 2 kinds of opinions regarding to the shape of it: quasi-cylindrical plume or curtain plume. Curtain plumes are  denser plasma sheet appearing as radial rays when observed edge on.\citet{gab09} have been expanded this curtain plume model with microplumes network. Quasi-cylindrical geometry of plumes have been more widely accepted. Many scholars have been investigated how plumes expand cylindrically. Some work have shown that the density structure of coronal holes or polar plumes follow a radial expansion \citep{woo99};while some others concluded a superadial expansion \citep{def97,fgu95}. However, there still lacks a quantitative plume shape model. This is what this paper wants to investigate. 

\section{Observations and Data analysis}

STEREO observatory was launched at Oct.25th 2006. Data collected to study the width of the plume's cross section in this paper are all taken by STEREO SECCHI with the wavelength of 171 \AA. I first assumed plumes have an expanding cylindrical tube shape with a circular cross section. Observations have shown that the light intensity of the central axis along a plume reduces as enhanced altitude. The light intensity variation across any plume of some height appears toughly as a gaussian curve. A cross section width or diameter was measured by full width half maximum(FWHM) principle at 4 different heights from the sun disk center: 1.04$R_s$,1.10$R_s$,1.16$R_s$ and 1.20$R_s$ and did this for total 31 plumes. Some measurements at 1.20$R_s$ are excluded if they are too fuzzy. 

Because of lacking enough statistics, a polynomial function with 4 parameters was assumed to approximate how the diameter of plumes vary as increasing height. These 4 unknown parameters of the polynomial function were calculated analytically from 4 measured average cross section diameters. Notice that this polynomial function model only describes lower part of plumes,particularly lower than 1.2 $R_s$ normally observed by EUV. The standard deviation of each calculated average diameter is also shown in Table 1. This standard deviation may be linked with the internal structure of plumes and can not represent purely as statistics error. Figure 1 shows how a width was measured by FWHM and where the plot of pixel value versus circular circumference was taken from the image. Figure \ref{fig:myfig3} contains graph of model results and data points. Obviously, data points are on the graph of the model itself.The final model obtained is as following,where $d$ is the plume diameter and $r$ is the height from the sun center:

\begin{equation}
d=20.5-55.5\,r+50.0\,r^2-14.9\,r^3
\end{equation}

\begin{deluxetable}{ccc}
\tablecolumns{3}
\tablewidth{0pc}
\tablecaption{Measured Plume widths and their standard deviations}
\startdata
\colhead{Plume height from the Sun center($R_s$)} & \colhead{Average value($R_s$)} & \colhead{standard deviation($R_s$)}\\
\tableline
 1.04$R_s$ & 0.0614516 & 0.0171694\\
 1.10$R_s$ & 0.0703548 & 0.0213363\\
 1.16$R_s$ & 0.0841613 & 0.0242818\\
 1.20$R_s$ & 0.0865333 & 0.0200067\\
\tableline
\enddata
\end{deluxetable}

\begin{figure}
\subfigure{
\includegraphics[height=60mm]{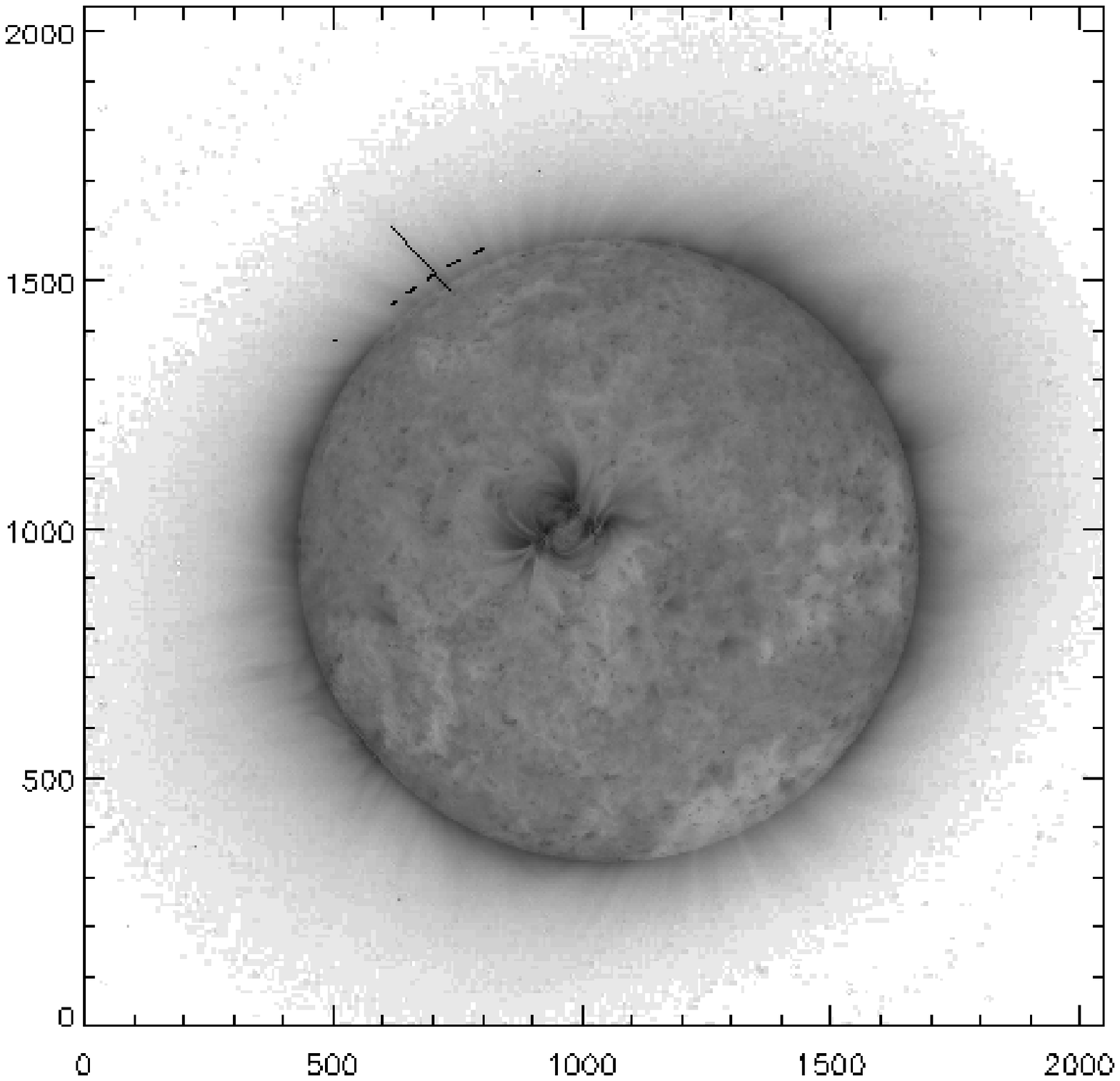}\label{fig:myfig1}
}
\subfigure{
\includegraphics[height=60mm]{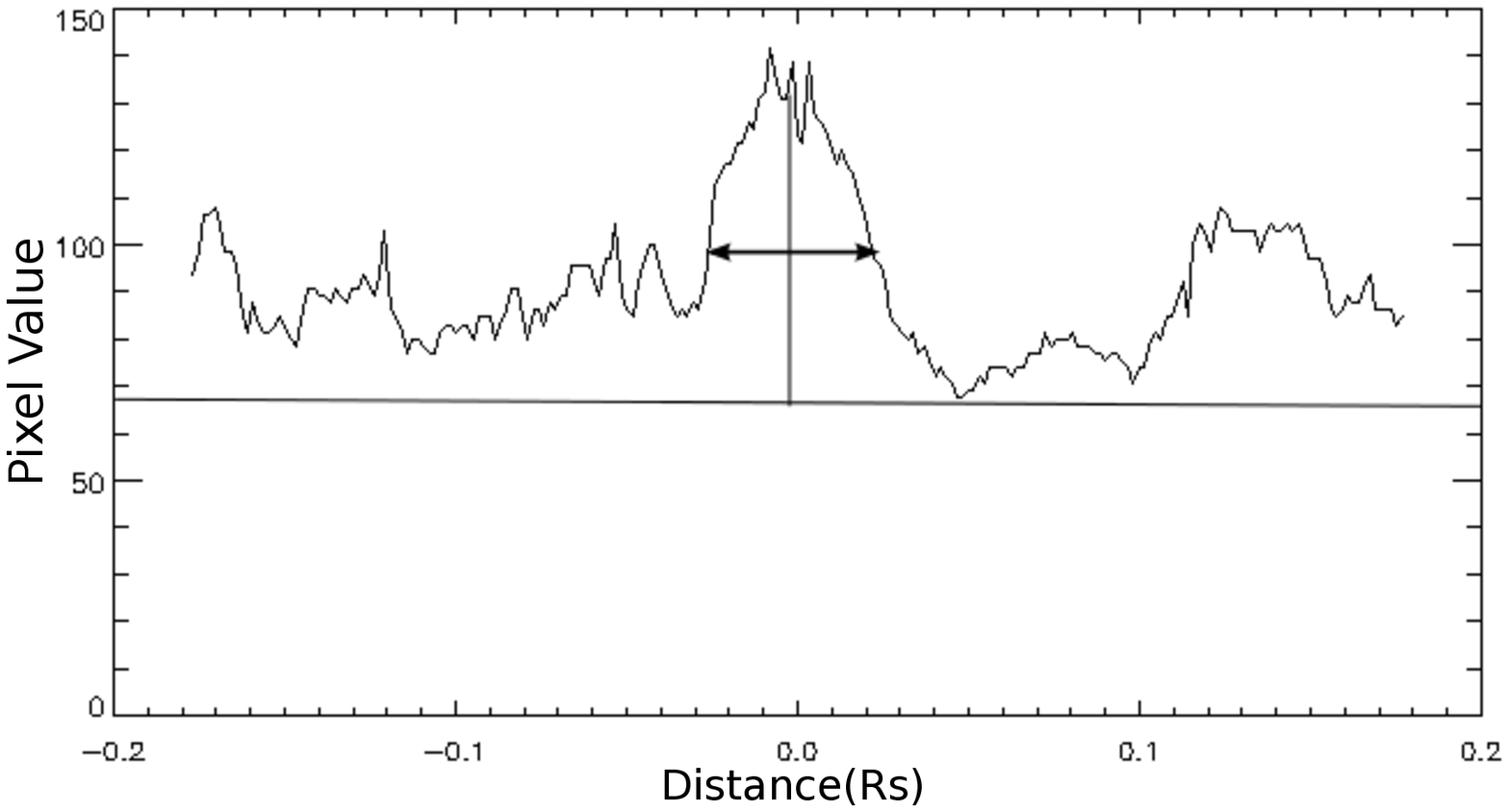} \label{fig:myfig2}
}
\caption{The figure on the left shows the line on the image of data taken;the right figure is the fructuation of  pixel value across the plume cross section \label{fig:myfig}}
\end{figure}
 
\begin{figure}
\includegraphics[height=60mm]{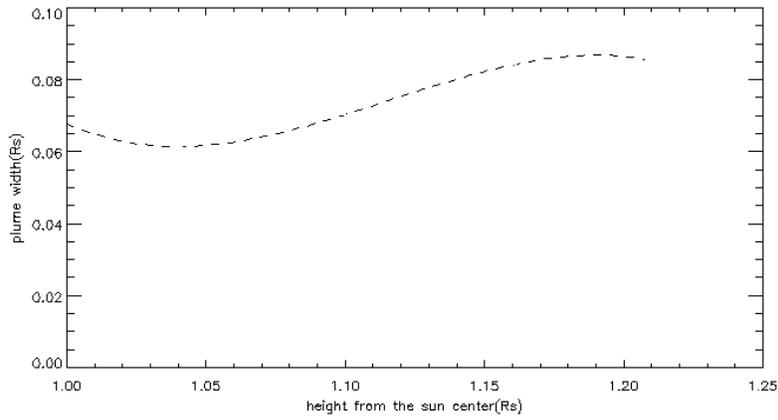}\label{fig:myfig3}
\caption{The model results is the solid curve with data represented by cross points  }
\end{figure}

\section{Conclusion and Discussion}

Diameters of each plume at different heights were measured by hand in this paper. Samples contain just 31 plumes considering the balance between accuracy and timing. If an automatic recognition computer program is used, large amount of samples can be analyzed to obtain good statistics.

\acknowledgments

I want to thank my previous advisor Jonathan Cirtain working in MSFC to initialize this project and funded me for 2 semesters.

Great thanks to faculties in Physics\&Astronomy Department in The University of Alabama and all my friends here. This work is never possible without all your help.

\end{document}